**Comparing Conventional and Deep Feature Models for Classifying Fundus Photography of Hemorrhages**

Tamoor Aziz, Chalie Charoenlarpnopparut, Srijidtra Mahapakulchai

School of Information, Computer and Communication Technology, Sirindhorn International Institute of Technology, Thammasat University, Pathum-Thani 12000, Thailand; tamoor.azi@dome.tu.ac.th;

* Correspondence: chalie@siit.tu.ac.th; Tel.: +66-2501-3505

**Abstract:**

Diabetic retinopathy is an eye-related pathology creating abnormalities and causing visual impairment, proper treatment of which requires identifying irregularities. This research uses a hemorrhage detection method and compares classification of conventional and deep features. Especially, method identifies hemorrhage connected with blood vessels or reside at retinal border and reported challenging. Initially, adaptive brightness adjustment and contrast enhancement rectify degraded images. Prospective locations of hemorrhages are estimated by a Gaussian matched filter, entropy thresholding, and morphological operation. Hemorrhages are segmented by a novel technique based on regional variance of intensities. Features are then extracted by conventional methods and deep models for training support vector machines, and results evaluated. Evaluation metrics for each model are promising, but findings suggest that comparatively, deep models are more effective than conventional features.

**Keywords:** features extraction; classification; image enhancement; hemorrhage detection; deep learning.

## 1. Introduction:

Diabetic retinopathy (DR) is a prevalent cause of vision loss among working-age adults. The statistics of DR patients have been projected to be 191 million by the year 2030 [1]. Initially, its diagnosis is almost impossible due to the absence of distinct symptoms. DR identification is crucial at the early phase because its timely treatment and medication may reduce the progression rate by 57% [2], approximately. Therefore, an annual examination is recommended for diabetes patients. Several surveys were conducted and highlighted that diabetes patients refused to have regular checkups because of lack of symptoms, time-consuming diagnostic process, and limited access to ophthalmologists [3]. DR falls into two main categories: non-proliferative diabetic retinopathy (NPDR) and proliferative diabetic retinopathy (PDR). NPDR weakens capillary walls and yields leakage of blood from vessels that compile microaneurysms (MAs). Later, ruptures turn MAs into hemorrhages (HEs). MAs and HEs are often term as red lesions. When the disease progresses then NPDR turns into PDR and angiogenic factors originate new blood vessels called neovascularization.

Eye experts use fluorescein angiography (FA), optical coherence tomography (OCT), and fundus photography for screening of DR [4]. FA is used to identify locations where blood vessels are closed or ruptured. OCT screening method provides a cross-sectional overview to determine the amount of fluid at retinal tissue and is used to evaluate the effectiveness of the adopted treatment. Next, fundus photography is an easy and immediate screening technique for documentation of DR progression and its improvement over time. Laser treatment, eye injections, or eye surgery can be recommended by an ophthalmologist in a case when DR is intimidating to eyesight [5]. Laser treatment helps to cure the neovascularization of blood vessels at the back of the eye. It stabilizes the changes that occur because of diabetes. Eye injection is used in the case of PDR to stop the emergence of new blood vessels. The benefit of this method is the improvement in eyesight. However, steroid injection produces excessive pressure inside the eye that may cause blood clots. Eye surgery is performed on an eye when a massive amount of blood accumulates at vitreous humour. Eye specialist removes some jelly-like substance that fills the space back of the eye.

Retinal fundus imaging is preferred for the initial screening phase because of its easy assessment and less expensive. Ophthalmologists capture retinal images using a fundus camera with an appropriate field of view (FOV). Early signs of DR are observed to determine its stage for medical prescription. Contrary to benefits, HEs detection is challenging due to certain impediments. Factors like blurriness and poor illumination may reduce diagnostic accuracy. Uneven lighting conditions may produce dark shades in retinal images and misleads detection. Blood vessels share intensity characteristics with HEs because of their similar appearance. Sometimes, HEs can be adjoined with blood vessels because they originate from them. Detection of those HEs is imperative for early screening of DR. HEs that reside at the retinal periphery are blended with the black background and are problematic to identify for computer-aided automatic detection. Appropriate selection of a deep network for classifying HEs is crucial to obtain promising results. Hence, these constraints cause HEs detection to be a challenging task. Fig. 1 shows the characteristics of fundus images.

The risk of human interpretation necessitates an efficient algorithm that can segment and classify hemorrhages effectively. The computer-based second interpreter expedites the diagnostic process and assists ophthalmologists in assessment. The proposed methodology addresses the problems of fundus images. A novel gradient-based adaptive gamma correction adjusts the brightness of fundus images adaptively. An automatic detection scheme is proposed by image calibration. The proposed smart-window-based adaptive thresholding (SWAT) segments the objects while isolating hemorrhages from blood vessels and the retinal periphery. Objects are classified based on the intuitive selection of conventional features by manipulating the visual appearance of hemorrhage in retinal fundus images. The statistical comparison of features for HEs classification using conventional and deep models is provided. This research paper uses various architectures of deep models to analyze which is suitable for HEs classification? Identification and detection of hemorrhages that resided at the retinal periphery and connected with blood vessels are the hallmarks of the proposed algorithm.

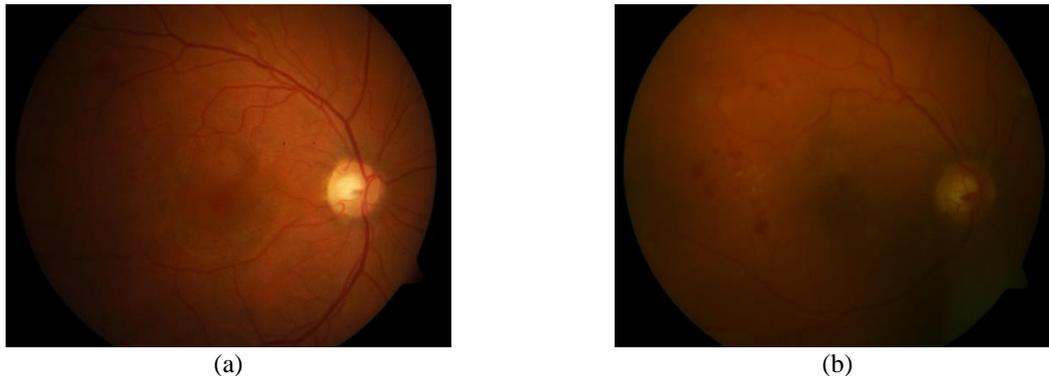

(a)          (b)

Fig. 1. Degraded retinal fundus images (a) uneven illumination (b) dark image because of low lighting condition.

## 2. Related Work:

N. Figueiredo et al. [6] proposed an algorithm to detect retinal abnormalities at the early stage of DR. This technique uses three classifiers for detection, including HEs. Novel features based on the inherent properties of lesions are used for classification. These features are extracted from wavelet bands, Hessian multiscale analysis, variational segmentation, and texture decomposition. The sensitivity and specificity of HEs detection are 86% and 90%, respectively. L. Tang et al. [7] propounded a splat feature classification method for HEs detection. The retinal image is partitioned into non-intersected segments called splats. The formation of each splat is based on similar color and spatial information. Shape, texture, the intersection of neighboring splats, and filter bank information are used. Later, optimum features are selected using the filter approach. This method achieves a 0.96 receiver operating characteristic curve (ROC). Detection of early signs of DR proposed by S. B. Junior and D. Welfer [8]. The technique is based on mathematical morphology to remove fovea and blood vessels because they share the intensity characteristics with HEs. This approach achieves 87.69% sensitivity and 92.44% specificity. The gradual elimination of blood vessel-based HEs detection technique is presented by L. Zhou et al. [9]. This technique deals with the HEs that are attached to the blood vessels by segmenting the dark regions, retinal vasculature, and HEs candidates. A binary image is manipulated further for providing good vascular connectivity and then removed gradually. Support vector machine (SVM) is trained using 49 features to classify candidates into non-HEs and HEs. The technique benchmarks promising results for two datasets. S. Karkuzhali and D. Manimegalai detect retinal abnormalities to classify fundus images into various DR stages [10]. Median filter, shade correction, Gaussian, and modified Kirsch filter are used to suppress noise and quality enhancement in preprocessing stage. The image is divided into non-overlapping patches of similar gray information. The Super-pixel method is applied to obtain the uneven grids. The gradient magnitude with toboggan segmentation is used for HEs segmentation. Feature vector and classifier mark images into various stages of DR.

The automatic segmentation of retinal lesions is presented by J. H. Tan et al. [11] using a novel single convolutional neural network (CNN). The proposed CNN model consists of 10-layers that classify retinal lesions simultaneously. The technique normalizes input images before network training. The proposed CNN model marks 0.6257 sensitivity on a large dataset. Another automatic detection of retinal lesions is proposed by C. Lam et al. [12]. The technique uses 1324 image patches for the training of the deep network. The sliding window method considers all the patches from the testing image to generate the probability map. This CNN model provides promising results for each type of lesion. A deep learning approach propounded by S. M. S. Islam et al. [13] for the detection and grading of DR. The technique focuses on early DR detection using a novel CNN network. The method is tested on a publicly available Kaggle dataset and reports a 0.851 quadratic weighted kappa score and 0.844 area under the curve. The technique for detection of red lesions using the You Look Only Once (YOLO-V3) algorithm is proposed by P. Pal et al. [14]. The contrast of the green channel is enhanced and then the bounding boxes of red lesions are obtained using the YOLO algorithm. Detection is performed using Darknet53 and logistic regression provides the confidence level of an object. The model is trained using Adam optimization and tested for red lesion detection. Objectness threshold is employed to reduce the false predictions. This technique scores 83.33% average precision. A synergy deep learning model is presented by K. Shankar et al. [15] to classify fundus images into DR stages. This technique removes noise from the edges in the preprocessing stage. Then, histogram-based segmentation obtains regions for the classification. Synergy deep learning model classifies images into severity levels. The algorithm is benchmarked on the Messidor dataset that shows promising results.

## 3. Method

This section provides a detailed explanation of the detection scheme. Fig. 2 shows the steps of the propounded HEs detection scheme. First, the image is preprocessed to enhance the quality. Then the prospective hemorrhage candidates are estimated. The objects are segmented using smart window-based adaptive thresholding. Finally, the objects are classified into hemorrhage and non-hemorrhage classes using features.

### 3.1 Dataset Description

The algorithm is trained and tested on the DIARETDB1 dataset [16]. The dataset contains 89 fundus images, of which five images are normal and the rest have various DR pathological symptoms. These images are captured by the 50-degree field of view using a fundus camera under different illumination conditions.

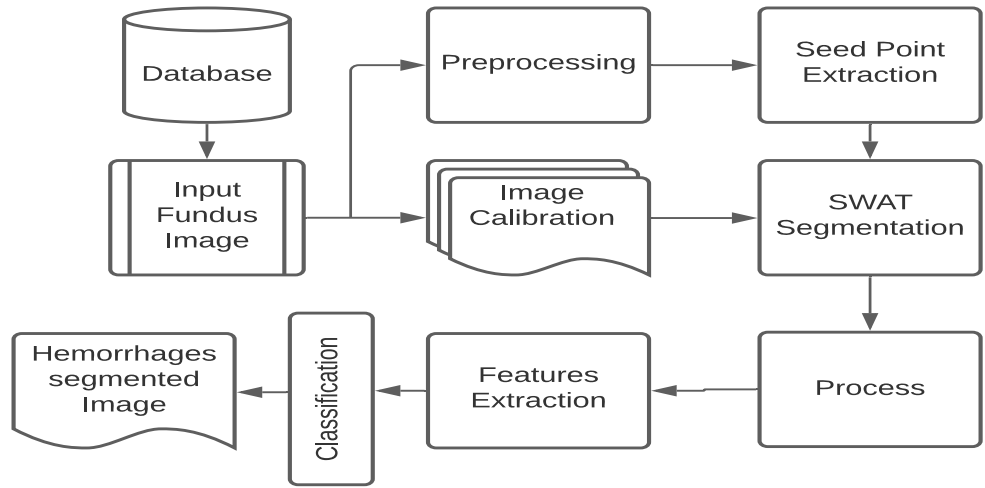

Fig. 2. Illustration of the proposed detection technique.

### 3.2 Preprocessing

Few images of the DIARETDB1 dataset have good brightness levels and contrast, while the majority of them are dark with low contrast. The quality of fundus images is enhanced using contrast limited adaptive histogram equalization (CLAHE) [17], gradient-based adaptive brightness adjustment (GAGC) [18], and non-linear unsharp masking [19]. CLAHE enhances contrast and reduces the effects of over-saturation by clipping intensity peaks. Our GAGC utilizes Sobel gradient information. Gamma correction [20] is applied using the adjusted threshold value of the Sobel operator. HEs can be attached to blood vessels and can only be separated when their regions are clearly defined. Therefore, fuzzy logic-based image sharpening using a non-linear filter is employed to sharpen the image. This method determines a fuzzy relationship between central and adjacent pixels in a $3 \times 3$ window. Sharpening filters work efficiently, but they introduce noise in the image. The non-linear property sharpens images and produces less noise than linear filters. The result of the preprocessing stage is provided in Fig. 3 (b).

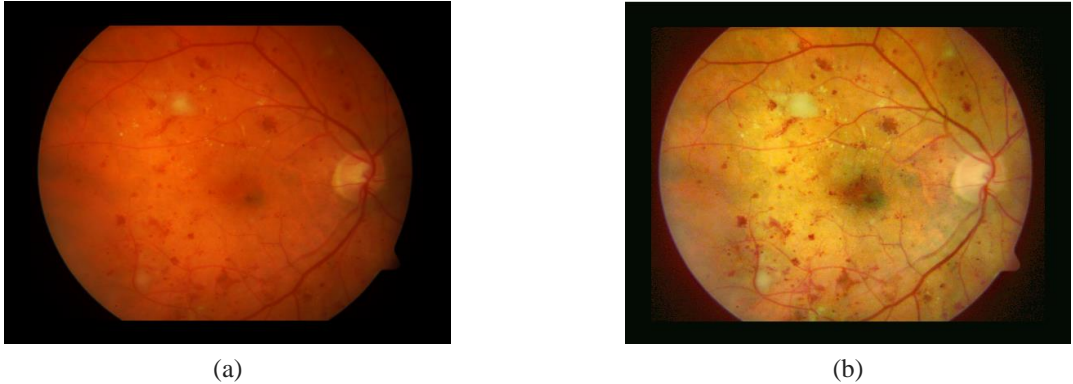

Fig. 3. Preprocessing of retinal fundus image (a) input fundus image (b) enhanced fundus image.

### 3.3 Seed Points Extraction

The detection process can be time-consuming if the entire image is considered for the search operation during detection. A good approach is to obtain prospective locations of objects to be detected and eliminate redundant information. This approach expedites detection with high accuracy. A similar technique manipulates the intensity profile of HEs in our work. HEs are dark objects surrounded by bright regions and share intensity characteristics with blood vessels and dark shades. This property suggests an inverted Gaussian matched filter [21], whose intensity values are low at the center and grow gradually beyond the center. This filter enhances HEs and blood vessels due to high correlation and yields low response wherever applied on the rest of the image, and can be depicted in Fig. 4 (a).

The redundant information is further reduced using the thresholding method. It depicted from the matched filtered image that low and high responses are close to each other. Therefore, entropy thresholding is employed [22] that eliminates unrequired information efficiently. This thresholding method finds cross entropies between quadrants of gray level co-occurrence matric (GLCM). The optimum threshold value from the gray range is selected successively, which minimizes the objective function. Fig. 4 (b) is a sample image of cross-entropy thresholding.

Elimination of blood vessels may also remove some of the HEs attached to them. Therefore, consideration of objects that correspond to blood vessels is imperative. The morphological opening is applied to break vasculature structure. This maneuver provides seed points for all types of HEs, including those are attached to blood vessels. Conversely, it increases the number of seed points for subsequent segmentation and classification stages and can be depicted in Fig. 4 (c).

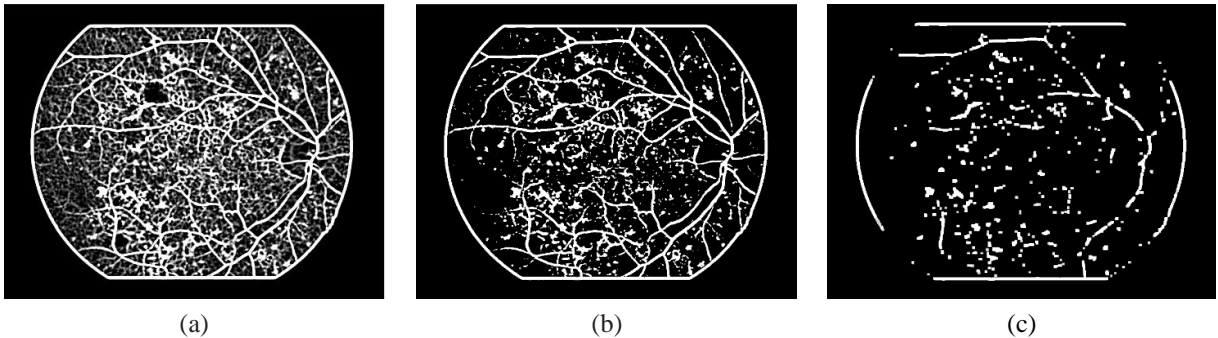

Fig. 4. Seed points extraction (a) response of matched filter (b) cross-entropy thresholding (c) morphological opened image.

### 3.4 Image Calibration

The HEs can be present at a jelly-like surface called the vitreous humour, and the black background does not contribute to the detection phase. A black background is darker than HEs and misleads the detection process. Therefore, it impedes the automatic detection of those HEs that reside at the retinal border. The black background is illuminated for effective and automatic detection. First, a median filter is applied on a green channel to suppress intensity variation at the background and then binarized. The resultant image is called the retinal mask that highlights the retinal area. Later, an eroded mask is subtracted from the retinal mask to get the retinal boundary. Calibrated image is obtained by adding an enhanced green channel, complemented retinal mask, and retinal border. A sample of the calibrated image can be depicted from Fig. 5 and is used for segmenting HEs.

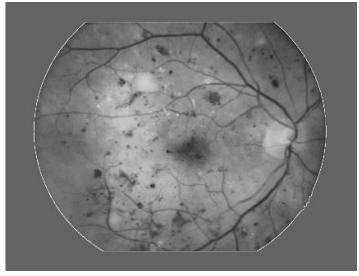

Fig. 5. Image calibration for feature extraction.

### 3.5 Smart Window-based Adaptive Thresholding Segmentation

A segmentation method is sensitive to the dissimilarity of objects and their surroundings, and dissimilarity can be in terms of intensities or textures. There are two challenges for segmenting HEs. First is a segmentation of HEs blended with the black background and located at the retinal rim. This background has been illuminated using image calibration. The second is a segmentation of HEs that are attached to blood vessels. Blood vessels and HEs share intensity characteristics and they are known as dark smooth regions. Therefore, a novel smart window-based adaptive thresholding (SWAT) is proposed that isolates HEs from blood vessels. This method is adaptive and segments HEs encompassed by various bright regions. A search space is defined for automatic detection to constrain segmentation within image range. Complemented binary mask, obtained in the previous section, is expanded 80 pixels wide to provide sufficient space for HEs residing at the retinal border.

Segmentation using a threshold value obtained by maximizing inter-region variance from image histogram [23]. This method determines the weighted variance $\sigma_B^2(j)$ between regions for a given threshold value $j$ as:

$$\sigma_B^2(j) = \sum_{z=1}^{R} \omega_z (\mu_z - \mu_T)^2 \qquad (1)$$

where $\mu_T$ is mean value of an original image, $\omega_z$ is the total probability of individual region $z$, and $\mu_z$ is the mean value of individual regions in $R = \{2,3,\ldots,19,20\}$ after thresholding. An optimum threshold value is taken successively by maximizing the inter-region variance as:

$$\sigma_B^2(\tau^*) = \max_{1 \leq j \leq L} \sigma_B^2(j), \quad j = \{0,1,2,\ldots,L-1\} \qquad (2)$$

Effectiveness $\eta$ of an optimum threshold $\tau^*$ depends upon a selection of an appropriate number of regions from $R$. An appropriate number of regions provides maximum effectiveness. $\eta$ is a ratio of weighted variance $\sigma_B^2(\tau^*)$ to the total variance $\sigma_T^2$ of image that can be calculated as:

$$\eta = \frac{\sigma_B^2(\tau^*)}{\sigma_T^2} \qquad (3)$$

SWAT initiates from seed points to segment retinal structures from the calibrated image, Fig. 5. The search process starts from the bounding box of a seed point. The calibrated image is cropped using the vertices $V = \{v_1, v_2, v_3, and\ v_4\}$ of a seed point. The cropped window $W_1(x,y)$ is thresholded iteratively until the appropriate number of regions from $R$ is selected as:

$$\vartheta = \begin{cases} R \rightarrow R+1, & if\ \eta < 0.8, \text{AND}\ R \leq 20 \\ \text{stop}, & \text{otherwise} \end{cases} \qquad (4)$$

where $\vartheta$ is a vector that contains $R-1$ threshold values. Equation 4 provides robustness in accordance with the regional diversity of HEs and foregrounds. In the case of bright foreground, fewer iterations are required to approach the stopping criteria that yield few numbers of regions. For the dark foreground, more iterations are required to reach $\eta \geq 0.8$, which requires more numbers of regions to perform effective segmentation. HEs are dark objects surrounded by various bright regions. The window is thresholded as:

$$W_2(x,y) = \begin{cases} 0, & if\ W_1(x,y) > \min(\vartheta) \\ 1, & else \end{cases} \qquad (5)$$

Where $\min(\vartheta)$ is the minimum threshold value of the vector $\vartheta$. There is a possibility that a window may have many HEs or dark objects after thresholding, so priority is given to the biggest ones because they are more dangerous for eyesight than the smaller HEs. Therefore, two large objects are kept and the rest are removed based on their area. This maneuver is applied such dark shades, often bigger sizes than HEs, cannot mislead the segmentation and actual HEs can be retained within the window. Furthermore, an object closer to the center of the window is more likely a HE than the other one. This probability criterion is proposed because seed points are extracted using the matched filter that models the intensity characteristic of HEs. Therefore, the object is eliminated using distance transform except one with minimum distance from the center of the window. The distance $d_i$ of the $i_{th}$ object from the center $W_2(x_c, y_c)$ of the window is calculated using:

$$d_i = \min\sqrt{\{W_2(x_c) - I_i(x)\}^2 + \{W_2(y_c) - I_i(y)\}^2} \qquad (6)$$

where $I_i(x)$ and $I_i(y)$ denote the $x$ and $y$ spatial locations of $i_{th}$ object, respectively, and $i = \{1,2\}$. The sizes of the HEs are bigger than the size of the window because they initiated from a seed point. The window must be expanded to capture the complete HEs using:

$$V = \begin{cases} v_1 \rightarrow v_1 - 5, & if\ q_1 = 1\ AND\ v_1 \cap S \\ v_2 \rightarrow v_2 - 5, & if\ q_2 = 1\ AND\ v_2 \cap S \\ v_3 \rightarrow v_3 + 10, & if\ q_3 = 1\ AND\ v_3 \cap S \\ v_4 \rightarrow v_4 + 10, & if\ q_4 = 1\ AND\ v_4 \cap S \end{cases} \qquad (7)$$

$Q = [q_1, q_2, q_3, q_4]$ contains information of border pixels. Binary variables $q_1, q_2, q_3,$ and $q_4$ correspond to left, top, right, and bottom border pixels, respectively. If all these variables are 0, then no further iteration is required because it shows the complete segmentation of the object. If any variable in $Q$ has a value of 1, it guarantees that the size of an object is bigger than the size of the window towards a particular direction. The window is expanded using Equation 7 and the calibrated image is cropped by the updated vector $V$.

The search space assists in performing segmentation automatically. Some of the seed points are redundant and belong to blood vessels and dark shades. A window may go beyond image range when segmenting blood vessels or dark shades. The condition on vector $S$ in Equation 7 determines whether vertices of vector $V$ lie within search space. Windows containing HEs and non-HEs objects are classified using features in the next section.

### 3.6 Features Extraction and Classification Stage

Support vector machine (SVM) is a statistical learning model used for classification by placing a hyperplane between positive and negative examples. Three sets of deep features were obtained from the hidden layers of VGG16 [24], ResNet50 [25], and AlexNet [26]. Four SVMs trained using conventional features and deep features to classify objects into HEs and non-HEs categories. Conventional features manipulate the visual appearance of HEs. For instance, HEs have sharp edges than macula, known as central vision. So, Laplacian-gradient features differentiate HEs from the macula. Blood vessels are line-shaped objects and HEs are comparatively circular objects. Therefore, connected component descriptors are useful to classify them. Color features help to distinguish dark shades from HEs. Opened or closed object's contour, number of corner points, and the spatial distance from the corners to the object's center are hand-crafted features. Hence, connected component [27], texture [28], color [29], and hand-crafted features are extracted to train SVM. While the VGG16, ResNet50, and AlexNet CNN models provide deep features for SVMs training.

## 4. Results, Comparison, and Discussion

The findings of the propounded detection scheme are reported in this section. Illustrations of performance metrics and the statistical comparison of various deep models are presented. The results can be pictorially be depicted in Fig. 6.

### 4.1 Data Preparation and Evaluation Metrics

The DIARETDB1 dataset is employed to detect HEs and compare various feature extraction models. This dataset is divided into training and testing subsets. The training subset is further separated into training and validation subsets. Windows obtained by the SWAT segmentation were annotated using ground truths. Twenty images are used to benchmark the performances of classifiers. The classification results are compared using sensitivity (SE) and specificity (SP) [30].

$$SE = \frac{TP}{TP + FN} \qquad (8)$$

$$SP = \frac{TN}{TN + FP} \qquad (9)$$

Where true-positive (TP) and true-negative (TN) are the truly predicted measurements by the classifier. TP is the rate of truly classified hemorrhage, while TN is the correct prediction rate of the negative class. Conversely, false-positive (FP) and false-negative (FN) are the measurements of the false predictions of the classifier. FP wrongly indicates that an object belongs to a hemorrhage, but actually, it does not. FN shows that hemorrhage is not present while the window contains a hemorrhage.

**4.2 Results**

Results represent that the false-negative (FN) rate of conventional features is higher than deep features. It states that conventional methods cannot identify some of HEs. Conversely, deep models are more capable of HEs identification. The classification results of deep and conventional models are provided in Table 1, while visually can be depicted from Fig. 6.

| Methods  | SE (%) | SP (%) |
|----------|--------|--------|
| SVM      | 88.98  | 97.67  |
| VGG16    | 95.88  | 94.87  |
| ResNet50 | 92.24  | 97.81  |
| AlexNet  | 92.21  | 98.24  |

Table 1. Comparison of the classification stage using various models on the DIARETDB1 dataset.

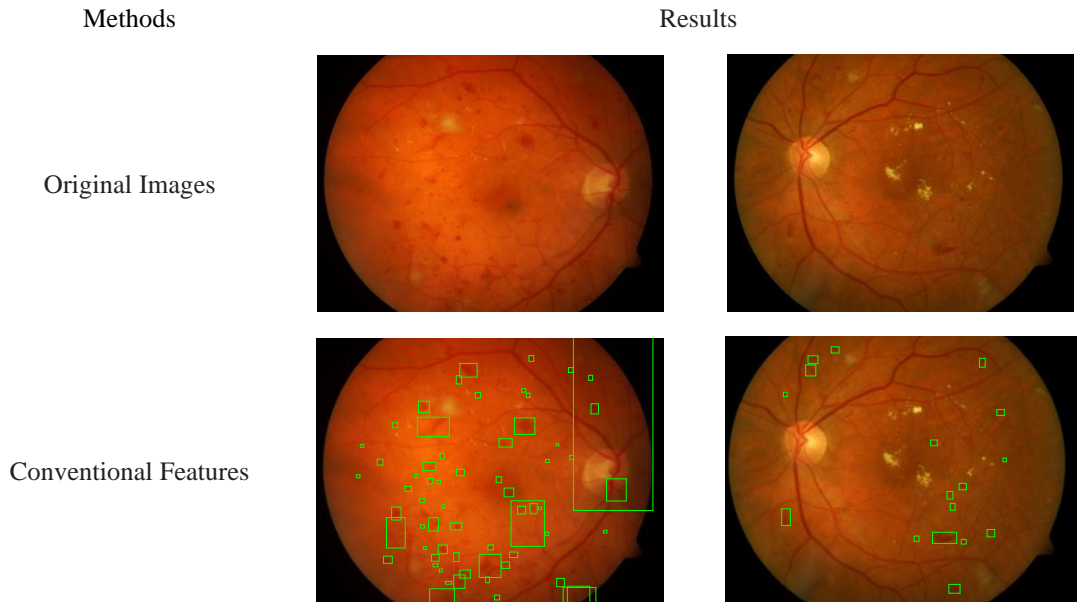

| Methods | Results |
|---------|---------|
| Original Images | |
| Conventional Features | |

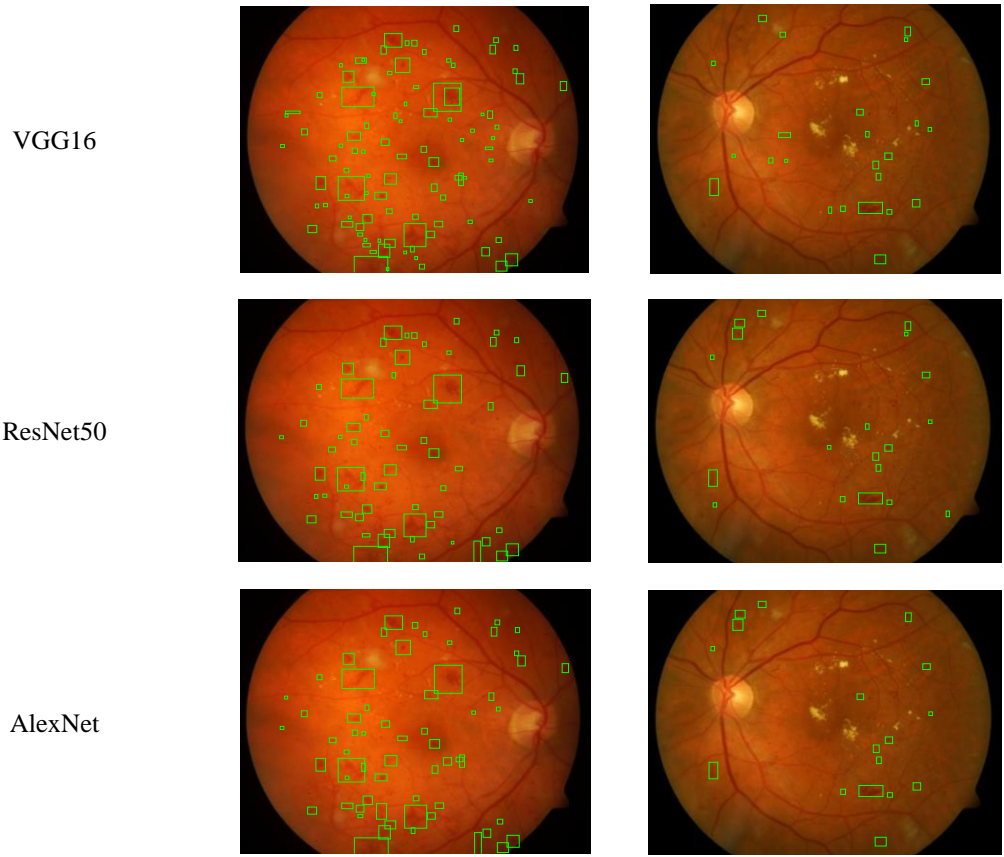

Fig. 6. Classification results (1st-row) Original Images (2nd-row) classification using conventional features (3rd-row) classification VGG16 (4th-row) classification using ResNet50 (5th-row) classification using AlexNet.

## 5. Discussion

The SE and SP observe the performances of the classification models. All the features extraction models show promising results and are applicable for the detection of DR. However, the FN rate of the SVM trained by the conventional features is the highest, resulting in a minimum SE than the other methods. The reason is obvious the small arteries of the blood vessels. Blood vessels are classified using connected components, but the small arteries also share these properties. Therefore, their similar appearance concerning the intensity and connected component characteristics misleads the classifier because they are labeled as a negative class. The conventional features of HEs and the arteries overlap; therefore, the highest misdetection rate. It marks 88.98% SE and 97. 67% SP. VGG16 CNN model has better SE that can detect more HEs than the conventional method. It marks the highest SE but less SP which states its false-positive (FP) rate is the highest compared to the other methods. The worst performance of VGG16 might be its high convergence rate towards the solution. The increased convergence rate has the drawback of oscillatory behavior around the optimum solution. Therefore, the network cannot converge to the optimum point for useful features. The SE and SP of this deep model are 95.88% and 94.87%, respectively. The performances of the ResNet50 and AlexNet are mediocre. They provide better SE than the convention method and better SP than the other two models. Evaluation metrics output 92.24% SE and 97.81% SP by the ResNet50. While the AlexNet also has a similar behavior for SE but the SP is considerably higher than ResNet50. Effectively, it yields the highest SP among all the methods. The statistics of AlexNet for SE and SP are 92.21% and 98.24%, respectively. Furthermore, the assessment of ResNet50 and AlexNet architectures reveals that the ResNet50 is unnecessarily deep. ResNet50 contains fifty layers, while AlexNet is eight layers deep. Therefore, AlexNet can be a good choice for HEs classification because it marks competitive results.

The assessment of the deep feature extraction models reveals that the arrangement of layers in deep models is crucial for a particular application. The increasing number of deep layers may not yield good results instead increases training time. For instance, AlexNet is shallower than VGG16 and ResNet50 but provides the highest SP of

98.24 and competitive SE of 92.21. While VGG16 is deeper than AlexNet and shallower than ResNet50 yields the highest SE of 95.88 and lowest SP of 94.87. ResNet50 is the deepest and marks mediocre results.

Furthermore, It is observed from the architectures of the pre-defined networks that the filter sizes of the first convolution layers of VGG16, ResNet50, and AlexNet are $3 \times 3$, $7 \times 7$, and $11 \times 11$, respectively. The small filter's size is appropriate for HEs classification because of its homogeneous property. HEs are regarded as dark smooth regions. VGG16 has the smallest filter size and identifies more HEs because its FN rate is the lowest among other deep models. Conversely, dark shades and small blood vessels mislead VGG16, resulting in the highest FP rate.

The analysis of the classification results recommends the deep model for the HEs classification. The reason could be that some conventional features may not be effective and mislead the classifier. The deep networks provide relevant features because they learn incrementally from the data. Therefore, no feature can mislead the classification stage. On the contrary, the CNN models take more time to learn from the windows and recognize them. They often need large numbers of training examples, depending upon the complexity of the data, for better performance. The conventional method needs comparatively less time and training examples to obtain the statistical features.

## 6. Conclusion

This research presents an automatic detection technique to compare various deep learning-based models with the conventional features extraction approach. The method first enhances the quality of the fundus images for a better appearance of pathological symptoms in the preprocessing stage. Then, the locations of the hemorrhages are estimated using seed points extraction that expedites the detection process. Deep and conventional features classify the objects into hemorrhages and non-hemorrhages. The research concept emerged from the problem highlighted by the research community that two types of hemorrhages are challenging to detect. First, the hemorrhages that are associated with the blood vessels. Second, the hemorrhages are located at the retinal border. Our detection scheme is suitable for all types, including those hemorrhages that reside at the vitreous humour. This study also prescribes that the deep features can better classify hemorrhages than the conventional methods; hence they are more efficient and suitable for the hemorrhages classification.

The assessment of performance metrics of deep modalities reveals that a shallow network produces competitive results compared to deep models. An intense deep network may not yield significant improvements but increases training time. In this study, AlexNet shows promising results despite the shallowest network. Therefore, a suitable network with its appropriate parameters is critical.

The research work's intuition is to present a fully-automated scheme for reducing the misdetection rate of hemorrhages by ophthalmologists interpreting fundus photographs. The method identifies hemorrhages in an interactive way that is easy to interpret for Diabetic retinopathy diagnosis. Furthermore, the locations of hemorrhages are highlighted, which might help the clinicians conclude the severity levels of the disease.

**Author Contributions:** TA: investigation, methodology, software implementation, writing original draft, conceptualization, and formal analysis.

CC: funding acquisition, project administration, supervision, writing—review, and editing.

All authors have read and agreed to the published version of the manuscript.

**Funding:** Thammasat University Research Fund, SIIT Research Fund.

**Data Availability Statement:** Publicly available dataset was analyzed in this study. This data can be found here: https://www.it.lut.fi/project/imageret/diaretdb1/index.html.

**Conflicts of Interest:** All authors in this paper have no potential conflict of interest.